\documentclass[showpacs,showkeys,11pt,
preprint,preprintnumbers,nofootinbib,
groupedaddress,superscriptaddress,amsmath,amssymb]{revtex4}
\usepackage{amsfonts}
\usepackage{graphics}
\usepackage{epsfig}
\usepackage{url}
\usepackage{multirow}
\usepackage{feynmp}
\newcommand {\be}{\begin{equation}}
\newcommand {\ee}{\end{equation}}
\newcommand {\ba}{\begin{eqnarray}}
\newcommand {\ea}{\end{eqnarray}}

\newcommand {\tanb}{$\tan\beta~$}
\newcommand {\ra}{\rightarrow}

\begin{document}
\title{Enhancement of Charged Higgs Production in Association With $W^{\pm}$ at Muon Colliders in General Two Higgs Doublet Models}
\author{M. Hashemi}
\email{hashemi_mj@shirazu.ac.ir}
\affiliation{Physics Department and Biruni Observatory, College of Sciences, Shiraz University, Shiraz 71454, Iran}

\begin{abstract}
The charged Higgs associated production with a $W^{\pm}$ boson has a smooth cross section as a function of the charged Higgs mass at muon colliders. The cross section in Minimal Supersymmetric Standard Model is about 25 $fb$ in the range 200 GeV $< m_{H^{\pm}} <$400 GeV with \tanb = 50. This is much larger than the corresponding cross section at an $e^{+}e^{-}$ collider which reaches a fraction of femtobarn. The observability of this charged Higgs production at a muon collider has been recently studied in an earlier work leading to the result that with 1 $ab^{-1}$, a 5 $\sigma$ signal can be observed throughout the aforementioned mass range. In this paper, results of a study based on a general two Higgs doublet model (type II and III) are presented and the cross section of the charged Higgs production in the most sensitive parameter space is evaluated. It is concluded that the cross section increases with increasing neutral Higgs boson masses involved in the s-channel diagram and can be as large as several picobarn with \tanb = 50. The region of ``physical Higgs boson mass'' parameter space which could lead to a 5 $\sigma$ signal at 50 $fb^{-1}$ is specified.  
\end{abstract}

\maketitle

\section{Introduction}
The Standard Model of particle physics contains one complex Higgs doublet which predicts a single neutral Higgs boson after electroweak symmetry breaking through the Higgs mechanism \cite{Higgs1,Higgs2,Higgs3,Higgs4,Higgs5}. Within SM, the radiative corrections arise the problem of the Higgs boson mass quadratic divergence as a function of the ultraviolet momentum cut-off ($\Lambda)$ used to regulate the loop integrals. These radiative corrections result in very large and un-natural values for the Higgs boson mass. There are theories which provide solutions to this problem such as supersymmetric theories which remove the Higgs boson mass divergence by introducing the so called ``super-partners'' for each particle being different by half unit of spin \cite{susy,susy2}. Such theories use a non-minimal Higgs sector. As an example, the Minimal Supersymmetric Standard Model (MSSM) requires two Higgs doublets to give masses to all leptons and quarks \cite{Martin}. Although the family of two Higgs doublet models (2HDM) is extensively used in different theories such as supersymmetry, it was originally introduced in \cite{lee} to describe the phenomenon of CP violation. In \cite{weinberg} it was shown that CP violation and Flavor Changing Neutral Currents (FCNC) can be naturally suppressed via the Natural Flavor Conservation (NFC) mechanism if a $Z_{2}$ symmetry is imposed on the Lagrangian. The most general potential using two Higgs doublets takes the form \cite{2hdm1,2hdm2}
\begin{align}
\mathcal{V} = & m_{11}^2\Phi_{1}^{\dag}\Phi_{1} + m_{22}^2\Phi_{2}^{\dag}\Phi_{2} - \left[m_{12}^2\Phi_{1}^{\dag}\Phi_{2}+\textnormal{h.c.}\right] \notag \\
&+\frac{1}{2}\lambda_1\left(\Phi_{1}^{\dag}\Phi_{1}\right)^2+\frac{1}{2}\lambda_2\left(\Phi_{2}^{\dag}\Phi_{2}\right)^2
+\lambda_3\left(\Phi_{1}^{\dag}\Phi_{1}\right)\left(\Phi_{2}^{\dag}\Phi_{2}\right)
+\lambda_4\left(\Phi_{1}^{\dag}\Phi_{2}\right)\left(\Phi_{2}^{\dag}\Phi_{1}\right) \notag \\
&+\left\lbrace\frac{1}{2}\lambda_5\left(\Phi_{1}^{\dag}\Phi_{2}\right)^2+\left[\lambda_6\left(\Phi_{1}^{\dag}\Phi_{1}\right)
+\lambda_7\left(\Phi_{2}^{\dag}\Phi_{2}\right)\right]\left(\Phi_{1}^{\dag}\Phi_{2}\right)+\textnormal{h.c.}\right\rbrace 
\end{align}
The $Z_{2}$ symmetry is thus defined as $\Phi_{i}\rightarrow(-1)^{i}\Phi_{i}$ ($i=1,2$) which has been discussed in details in \cite{2hdm3}. The general potential mentioned above contains the following free parameters
\be
\lambda_1,\lambda_2,\lambda_3,\lambda_4,\lambda_5,\lambda_6,\lambda_7,m_{12}^{2},\tan\beta 
\ee
in the general basis or 
\be
m_h,m_H,m_A,m_{H^{\pm}},\sin(\beta-\alpha),\lambda_6,\lambda_7,m_{12}^2,\tan\beta
\ee
in the physical Higgs masses basis.
The parameters $\lambda_{5-7}$ and $m_{12}^{2}$ can in general be complex leading to $CP$-violation effects which are avoided in this work. Therefore all parameters are assumed to be real. Since 2HDM's suffers from the same quadratic divergences of the Higgs boson masses, a combination of supersymmetry and 2HDM might be a reasonable idea to avoid such problems. The MSSM is one of such models which belongs to 2HDM family. It is, however, a very constrained model leaving only two free parameters taken usually as $m_{A}$ and \tanb (the ratio of vacuum expectation values of the two Higgs fields $\Phi_1$ and $\Phi_2$). In this model $\lambda_i$ can be expressed in terms of electroweak gauge couplings as follows \cite{mm1,mm2,mm3}:
\begin{align}
\lambda_1=\lambda_2=\frac{g^2+g^{\prime 2}}{4},&  \notag \\
\lambda_3=\frac{g^2-g^{\prime 2}}{4},&  \notag \\
\lambda_4=-\frac{g^2}{2},&  \notag \\
\lambda_5=\lambda_6=\lambda_7=0,&  \notag \\
m_{12}^{2}=m_A^2\cos\beta\sin\beta.&  
\label{lambdaseq}
\end{align}
As is seen from Eq. \ref{lambdaseq}, in this model $\lambda_6=\lambda_7=0$ which is a requirement to respect the $Z_2$ symmetry and avoid $CP$-violation and FCNC at tree level. Moreover, $\lambda_5$ is also zero. This is a characteristic feature of SUSY models \cite{HHH,HHH2}. Therefore, in order to respect SUSY and be close to MSSM, the following setting is adopted throughout the paper:
\be
\lambda_5=\lambda_6=\lambda_7=0
\ee
This setting relates $m_{12}^2$ and $m_A$ in the same way as in MSSM, through the general 2HDM relation
\be
m_A^2=\frac{m_{12}^2}{\sin\beta\cos\beta}-\frac{v^2}{2}(2\lambda_5+\lambda_6\cot\beta+\lambda_7\tan\beta)
\ee
Therefore, working in the ``physical Higgs masses'' basis, the phase space is reduced to contain the following subset of free parameters:
\be
m_h,m_H,m_A,m_{H^{\pm}},\sin(\beta-\alpha),\tan\beta.
\ee
This scenario leaves the Higgs boson masses free contrary to the case of MSSM in which all Higgs boson masses can be related to $m_A$ at tree level. The Higgs bosons freedom appeared in the general 2HDM mentioned above can lead to enhancement of cross section of some processes such as the charged Higgs associated production with a W boson.\\

\section{The Charged Higgs Mass Limits: Results from Direct and Indirect Searches}
The charged Higgs boson, due to being charged, is a signature of models with at least two Higgs doublets. There has been continuous searches for this particle in previous and current experiments. The LEP II experiment excluded a charged Higgs with $m_{H^{\pm}}<89~ \textnormal{GeV}$ for all \tanb assumptions \cite{lepexclusion1}. The indirect searches which use constraints on the neutral Higgs bosons masses to set limits on the charged Higgs mass exclude a charged Higgs with $m_{H^{\pm}}<125~ \textnormal{GeV}$ \cite{lepexclusion2}. \\
The Tevatron results from the D0 Collaboration \cite{d01,d03,d04} and the CDF Collaboration \cite{cdf1,cdf4} allow $2 <$ tan$\beta < 30$ for $m(H^{\pm}) > 80 $ GeV while more $\tan \beta$ values are available for higher charged Higgs masses. The most recent results on the charged Higgs searches come from ATLAS and CMS experiments at LHC \cite{atlasdirect,cmsdirect}. They use integrated luminosities of 4.6 and 2.3 $fb^{-1}$ respectively and both indicate that a light charged Higgs with $m_{H^{\pm}}< 140 ~\textnormal{GeV}$ is excluded with \tanb $>~ 10$. \\
The B-Physics constraints provide the strongest limits on the charged Higgs mass in 2HDM Type II and III. The study of the $b\ra s\gamma$ transition process using CLEO data excludes a charged Higgs mass below 295 GeV at 95 $\%$ C.L. in 2HDM Type II with \tanb higher than 2 \cite{B1}. A combination of all flavor data shows that the available part of the ($m_{H^{\pm}},\tan\beta$) plane is limited to $300<m_{H^{\pm}}<800$ GeV and $1<\tan\beta<70$ \cite{B11}. Therefore the indirect limits from flavor data favors a heavy charged Higgs. Including that limit makes charged Higgs searches at LHC hopeless. In fact putting all together, including LHC potential discovery contours \cite{CMScontour} and the indirect limits from flavor data shows that a charged Higgs with $m_{H^{\pm}}<600$ GeV may be out of reach by CMS and ATLAS experiments \cite{B111}. \\

\section{Charged Higgs at Future Linear Colliders}
The LHC experiment has special characteristics: the hadronic environment of $pp$ collisions, large final state particle multiplicity and the large background cross sections. In addition the charged Higgs signal cross section decreases at heavy mass region leaving few events for some charged Higgs masses. This situation motivates a linear collider with leptonic input beams such as International Linear Collider (ILC) or Compact Linear Collider (CLIC) \cite{ilc,rdr,clic,clic_cdr} with $e^{+}e^{-}$ as the input beam or the Muon Collider at Fermilab \cite{m1,m2,m3}. The center of mass energy is expected to be 0.5 TeV at ILC and 0.5 to 3 TeV at CLIC while the muon collider is expected to operate at 0.5 and 1 TeV. \\
The linear collider studies provide positive prospects for a heavy charged Higgs boson search. The $e^{+}e^{-}$ collider studies results can be found in \cite{pair,sch1,sch2,sch3,sch6,sch7}. The discovery potential of these analyses is limited to the region below $\sqrt{s}/2$ or slightly above if off-shell effects are included. In a work presented in \cite{mumuWH} it was shown that the cross section of $\mu^+\mu^- \ra W^{\pm}H^{\mp}$ is almost independent of the charged Higgs mass in the range 200 GeV $ < m_{H^{\pm}}< $ 400 GeV and therefore may provide opportunity to search for a charged Higgs in the aforementioned mass range. At linear $e^{+}e^{-}$ colliders, this channel has a very small cross section as studied in \cite{WH1,WH2} followed by studies of possible enhancement of the cross section by including quark and Higgs-loop effects in \cite{WH3}. In \cite{WH4,WH5} it was concluded that only few events of this kind may be observed at $e^{+}e^{-}$ linear colliders. 
\section{Associated Production of Charged Higgs Boson at a Muon Collider}
\begin{figure}
\begin{minipage}[t]{0.4\linewidth}
\centering
\unitlength=1mm
\begin{fmffile}{schannel}
\begin{fmfgraph*}(40,25)
\fmfleft{i1,i2}
\fmfright{o1,o2}
\fmflabel{$\mu^-$}{i1}
\fmflabel{$\mu^+$}{i2}
\fmflabel{$H^+$}{o2}
\fmflabel{$W^-$}{o1}
\fmf{fermion}{i1,v1,i2}
\fmf{dashes}{v2,o2}
\fmf{photon}{v2,o1}
\fmf{dashes,label=$h,,H,,A$}{v1,v2}
\end{fmfgraph*}
\end{fmffile}
\end{minipage}
\hspace{0.5cm}
\begin{minipage}[t]{0.4\linewidth}
\centering
\unitlength=1mm
\begin{fmffile}{tchannel}
\begin{fmfgraph*}(40,25)
\fmfleft{i1,i2}
\fmfright{o1,o2}
\fmflabel{$\mu^-$}{i1}
\fmflabel{$\mu^+$}{i2}
\fmflabel{$H^+$}{o2}
\fmflabel{$W^-$}{o1}
\fmf{fermion}{i1,v1}
\fmf{fermion}{v2,i2}
\fmf{fermion,label=$\nu_{\mu}$}{v1,v2}
\fmf{dashes}{v2,o2}
\fmf{photon}{v1,o1}
\end{fmfgraph*}
\end{fmffile}
\end{minipage}
\caption{The $s-$channel (left) and $t-$channel (right) diagrams involved in the signal process. \label{diagrams}}
\end{figure}
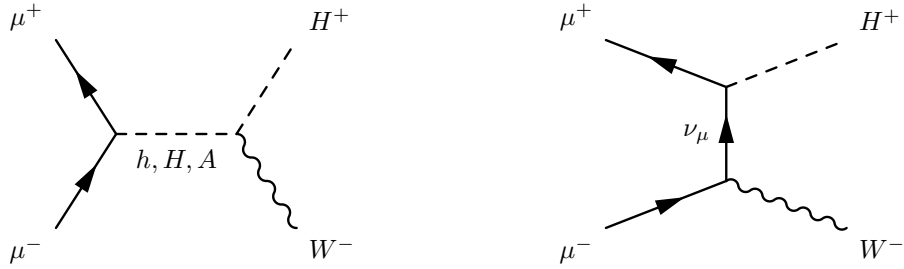

As stated in the previous section, the charged Higgs production in association with $W^{\pm}$ at a muon collider has a very larger cross section compared to the same process in $e^+e^-$ colliders due to the enhancement of the Yukawa couplings of the neutral and charged Higgs bosons involved in the s-channel and t-channel diagrams as shown in Fig. \ref{diagrams}. In a recent work \cite{mywh}, this process was studied in the framework of MSSM and it was shown that looking at $H^{\pm} \ra t\bar{b}$, the signal has a discovery potential at about 1 $ab^{-1}$. This needs a large amount of data, although the discovery potential is the same in the mass range 200 GeV $< m_{H^{\pm}} <$ 400 GeV. In addition to MSSM scenario, a similar study of this channel in \cite{mumuWH2} showed that in a general 2HDM with CP-violating terms, the signal cross section could be much higher than that in MSSM. In this study a general 2HDM without $CP$-violation is taken as the theoretical model and the effect of heavy neutral Higgs bosons in the production process is studied. The details of the study will follow in the next section. 
\section{The $W^{\pm}H^{\mp}$ Cross Section in a General 2HDM}
In this section the cross section of $\mu^+\mu^-\ra H^\pm W^\mp$ is studied in a general 2HDM Type II and III. As stated in the introduction, supersymmetry is also assumed. Therefore the theoretical model is a supersymmetric 2HDM of the type II and/or III.
The heavy charged Higgs boson predominantly decays to a pair of $t\bar{b}$ within MSSM and the reconstruction of this decay channel is easier than the decay to $\tau\bar{\nu}$ due to the existence of the missing energy. In a general 2HDM the coupling of charged Higgs boson and the $W^{\pm}$ depends on the choice of $\alpha$ and $\beta$ parameters as well as the neutral Higgs boson in the vertex:
\be
HH^+W^-: \sin(\beta-\alpha) ,~~~~~~~~~~ hH^+W^-: \cos(\beta-\alpha) .
\label{HVcoupling}
\ee
The above couplings are independent of the type of 2HDM. If $\cos(\beta-\alpha)\simeq 0$, the setting which corresponds to the decoupling limit \cite{2hdm_decoupling}, the charged Higgs only couples to the heavy neutral $CP$-even Higgs boson and the charged Higgs decay to $t\bar{b}$ is dominant up to the threshold of $m_H+m_W$. On the contrary, with $\sin(\beta-\alpha)\simeq 0$ (the non-decoupling limit), the charged Higgs decay to $W^{\pm}h^0$ competes the decay to $t\bar{b}$. Since the charged Higgs decay to $t\bar{b}$ is the topic of this study, a special care has to be taken for that. First notice that the four types of 2HDM's are expressed in terms of different couplings of the Higgs bosons with leptons and quarks in the Yukawa sector as follows \cite{yukawa}:
\begin{align}
-\mathcal{L}=&\frac{1}{\sqrt{2}}\bar{D}\left \lbrace\kappa^D s_{\beta-\alpha}+\rho^D c_{\beta-\alpha} \right \rbrace Dh
+\frac{1}{\sqrt{2}}\bar{D}\left \lbrace\kappa^D c_{\beta-\alpha}-\rho^D s_{\beta-\alpha} \right \rbrace DH + \frac{i}{\sqrt{2}}\bar{D}\gamma_5\rho^D DA \notag\\ 
&+\frac{1}{\sqrt{2}}\bar{U}\left \lbrace\kappa^U s_{\beta-\alpha}+\rho^U c_{\beta-\alpha} \right \rbrace Uh
+\frac{1}{\sqrt{2}}\bar{U}\left \lbrace\kappa^U c_{\beta-\alpha}-\rho^U s_{\beta-\alpha} \right \rbrace UH - \frac{i}{\sqrt{2}}\bar{U}\gamma_5\rho^U UA \notag\\
&+\frac{1}{\sqrt{2}}\bar{L}\left \lbrace\kappa^L s_{\beta-\alpha}+\rho^L c_{\beta-\alpha} \right \rbrace Lh
+\frac{1}{\sqrt{2}}\bar{L}\left \lbrace\kappa^L c_{\beta-\alpha}-\rho^L s_{\beta-\alpha} \right \rbrace LH + \frac{i}{\sqrt{2}}\bar{L}\gamma_5\rho^L LA \notag\\
&+\left[\bar{U}\left \lbrace V_{CKM}\rho^D P_R-\rho^U V_{CKM} P_L \right \rbrace DH^+ + \bar{\nu}\rho^LP_RLH^+ + \textnormal{h.c.} \right],
\label{Lagrangianeq}
\end{align}
where $s_{\beta-\alpha}=\sin(\beta-\alpha)$, $c_{\beta-\alpha}=\cos(\beta-\alpha)$, $\rho^Q=\lambda^Q\kappa^Q$, $\kappa^Q=\sqrt{2}\frac{m^Q}{v}$, and $\lambda^Q$ determine the type of the 2HDM \cite{types} according to Tab. \ref{tabtypes}:
\begin{table}[h]
\begin{tabular}{|c|c|c|c|c|}
\hline
\multicolumn{5}{|c|}{Type}\\
& I & II & III & IV \\
\hline
$\rho^D$ & $\kappa^D \cot\beta$ & $-\kappa^D \tan\beta$ & $-\kappa^D \tan\beta$ & $\kappa^D \cot\beta$ \\
\hline  
$\rho^U$ & $\kappa^U \cot\beta$ & $\kappa^U \cot\beta$ & $\kappa^U \cot\beta$ & $\kappa^U \cot\beta$ \\
\hline  
$\rho^L$ & $\kappa^L \cot\beta$ & $-\kappa^L \tan\beta$ & $\kappa^L \cot\beta$ & $-\kappa^L \tan\beta$ \\
\hline  
\end{tabular}
\caption{The four types of a general 2HDM in terms of the couplings in the Higgs-fermion Yukawa sector. \label{tabtypes}}
\end{table}
As seen from Tab. \ref{tabtypes}, the $H^{\pm}\ra t\bar{b}$ is suppressed at high \tanb in 2HDM type I and IV as the coupling is proportional to $\cot\beta$. Therefore the analysis is confined within a 2HDM type II and III. \\
At a muon collider the t-channel diagram contribution shown in Fig. \ref{diagrams} is small due to the small $H^{\pm} \ra \mu\bar{\nu}$ coupling. The s-channel diagrams are dominant since heavy neutral Higgs bosons are involved in the propagator. With $s_{\beta-\alpha}=1$ (the decoupling limit), $hH^+ W^-$ coupling vanishes and only  $A$ and $H$ contribute to the diagram. Since an SM-like theory favors a light $h$, one can choose $m_h\simeq 100$ GeV, thus the parameter space contains $m_{H^{\pm}}$, $m_A$ and $m_H$ as free parameters for fixed values of $\alpha$ and $\beta$ satisfying $s_{\beta-\alpha}=1$. The decoupling limit is a scenario with a light SM-like $h$ and heavy Higgs bosons with $m_{H^{\pm}}\simeq m_H \simeq m_A$. The charged Higgs mass upper limit is $m_{H^{\pm}} \simeq 420$ GeV for $\mu^+\mu^-\ra H^\pm W^\mp$ with a collider with $\sqrt{s}=500$ GeV. However, the contribution of the neutral Higgs bosons in the s-channel diagram starts to be sizable when their hypothetical masses exceeds $m_{A/H}\simeq 450$ GeV as evaluated by CompHep 4.5.1 \cite{comphep,comphep2}. This is a region which arises problems with the Higgs potential tree level unitarity and perturbativity as checked by Two Higgs Doublet Model Calculator (2HDMC 1.1) \cite{2hdmc}. In this region a problem with $\delta_{\rho}$ also appears as is described. In 2HDM, the relation $\rho_0=M^2_W/M^2_Z\cos^2 \theta_{W} = 1$ holds at tree level. At the one loop level there may be deviations from this equation which is written in the form $\rho=\rho_0+\delta_{\rho}$. For a 2HDM, $\delta_{\rho}$ is expressed in terms of the Higgs bosons masses as \cite{deltarho}
\begin{align}
\delta \rho _{\textnormal{2HDM}} = & \frac{G_F}{8\sqrt{2}\pi^2} \{  
M^2_{H^{\pm}}\left[1-\frac{M^2_{A}}{M^2_{H^{\pm}}-M^2_{A}}
\ln\frac{M^2_{H^{\pm}}}{M^2_A}\right] \notag \\ &+c^2_{\beta-\alpha}M^2_{h}\left[\frac{M^2_A}{M^2_A-M^2_h}\ln\frac{M^2_A}{M^2_h}-\frac{M^2_{H^{\pm}}}{M^2_{H^{pm}}-M^2_h}
\ln\frac{M^2_{H^{\pm}}}{M^2_h}\right] \notag \\
&+s^2_{\beta-\alpha}M^2_{H}\left[\frac{M^2_A}{M^2_A-M^2_H}\ln\frac{M^2_A}{M^2_H}-\frac{M^2_{H^{\pm}}}{M^2_{H^{pm}}-M^2_H}
\ln\frac{M^2_{H^{\pm}}}{M^2_H}\right] \}
\end{align}
With degenerate masses $M_A\simeq M_H \simeq M_{H^{\pm}}$, one obtains $\delta_{\rho}\ra 0$, however if the mass splitting between the Higgs bosons is large, $\delta_{\rho}$ deviates from 0 and obtains large values. Therefore the case $s_{\beta-\alpha}=1$ is not suitable if large cross sections for $\mu^+\mu^-\ra H^\pm W^\mp$ are explored in the mass regions defined by $m_{H^{\pm}}<420$ GeV (required by kinematic threshold) and $m_A\simeq m_H > 450$ GeV.\\
On the contrary, the case $s_{\beta-\alpha}=0$ (corresponding to the non-decoupling limit) can be understood from the Lagrangian Eq. \ref{Lagrangianeq}, which is summarized into the following forms for the type II and type III 2HDM's:
\begin{align}
-\mathcal{L}_{\textnormal{type II}}= & -\bar{D}\tan\beta \frac{m_D}{v} Dh+\bar{D} \frac{m_D}{v} DH-i\bar{D}\gamma_5\tan\beta \frac{m_D}{v} DA \notag\\
& +\bar{U} \cot \beta \frac{m_U}{v} Uh +\bar{U} \frac{m_U}{v} UH - i \bar{U} \gamma_5 \cot \beta \frac{m_U}{v} UA \notag\\
&-\bar{L}\tan\beta \frac{m_L}{v} Lh+\bar{L} \frac{m_L}{v} LH-i\bar{L}\gamma_5\tan\beta \frac{m_L}{v} LA,
\label{Lagrangianeq2}
\end{align}
\begin{align}
-\mathcal{L}_{\textnormal{type III}}=&-\bar{D}\tan\beta \frac{m_D}{v} Dh+\bar{D} \frac{m_D}{v} DH-i\bar{D}\gamma_5\tan\beta \frac{m_D}{v} DA \notag \\
&+\bar{U}\cot\beta \frac{m_U}{v} Uh+\bar{U} \frac{m_U}{v} UH-i\bar{U}\gamma_5\cot\beta \frac{m_U}{v} UA \notag \\
&+\bar{L}\cot\beta \frac{m_L}{v} Lh+\bar{L} \frac{m_L}{v} LH+i\bar{L}\gamma_5\cot\beta \frac{m_L}{v} LA.
\label{Lagrangianeq3}
\end{align}
Both Eqs. \ref{Lagrangianeq2} and \ref{Lagrangianeq3} show that with $s_{\beta-\alpha}=0$, the heavier neutral $CP$-even Higgs boson is SM-like with a coupling to fermions ($f$) proportional to $\frac{m_f}{v}$. This particle denoted by $H$ can be light enough to appear like the signal observed around 125 GeV by the ATLAS and CMS collaborations \cite{125atlas,125cms} while the lightest neutral Higgs boson may have enhanced or suppressed couplings with fermions depending on the type of 2HDM and its production rate may deviate from SM. The main production process for the neutral Higgs bosons at LHC is $gg\ra h$ which proceeds through a $t$ and $b$ quark triangular loop. The top quark, being heavier than the $b$ quark, has a stronger coupling with $h_{SM}$, and the SM production process proceeds mainly via a top quark loop. In a 2HDM type II with $s_{\beta-\alpha}=0$, the top quark coupling with $h$ is suppressed by $\cot \beta$, while the $b$ quark induced loop is enhanced by $\tan \beta$. On the other hand the branching ratio of decays, $h\ra b\bar{b}$ and $h\ra \gamma\gamma$ (which involves a fermion loop) will be enhanced (suppressed) by $\tan^2 \beta$ ($\cot^2 \beta$) in a 2HDM type II (III) with $s_{\beta-\alpha}=0$. A judgment on the total production rate including decays thus awaits a knowledge of the value of $\tan \beta$.\\
\section{Results}
\begin{figure}
\begin{center}
\includegraphics[width=0.7\textwidth]{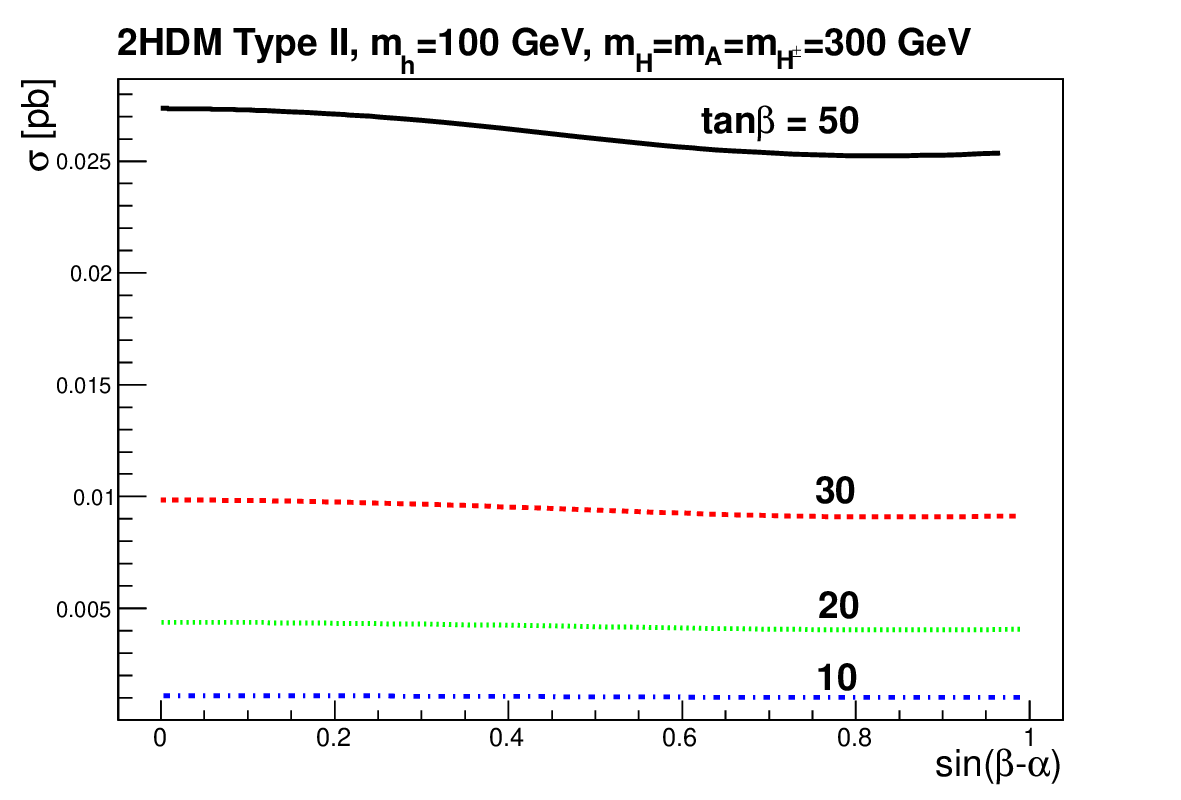}
\end{center}
\caption{The $\alpha$ scan of the cross section of the signal for fixed values of \tanb.}
\label{alphascan}
\end{figure}

\begin{figure}
\begin{center}
\includegraphics[width=0.8\textwidth]{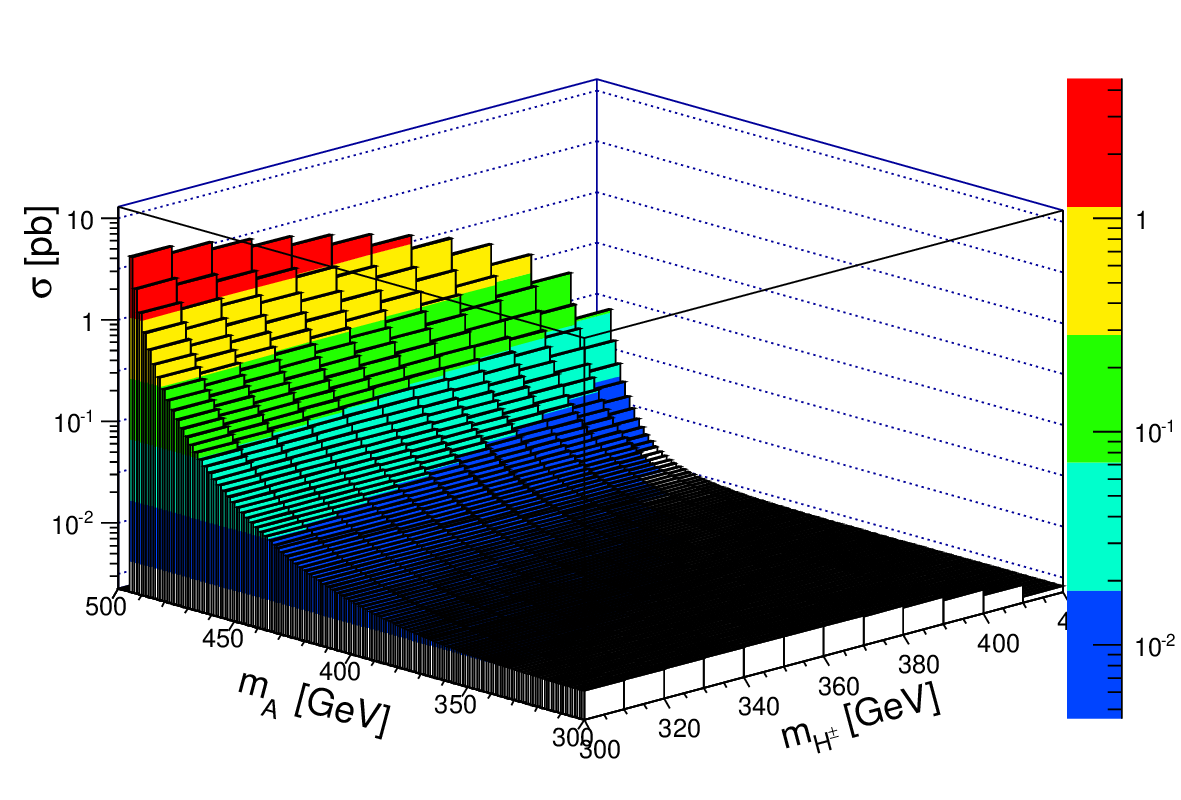}
\end{center}
\caption{The signal cross section in 2HDM type II as a function of $m_{H^{\pm}}$ and $m_A$ with \tanb= 20.}
\label{xsec20}
\end{figure}

\begin{figure}
\begin{center}
\includegraphics[width=0.8\textwidth]{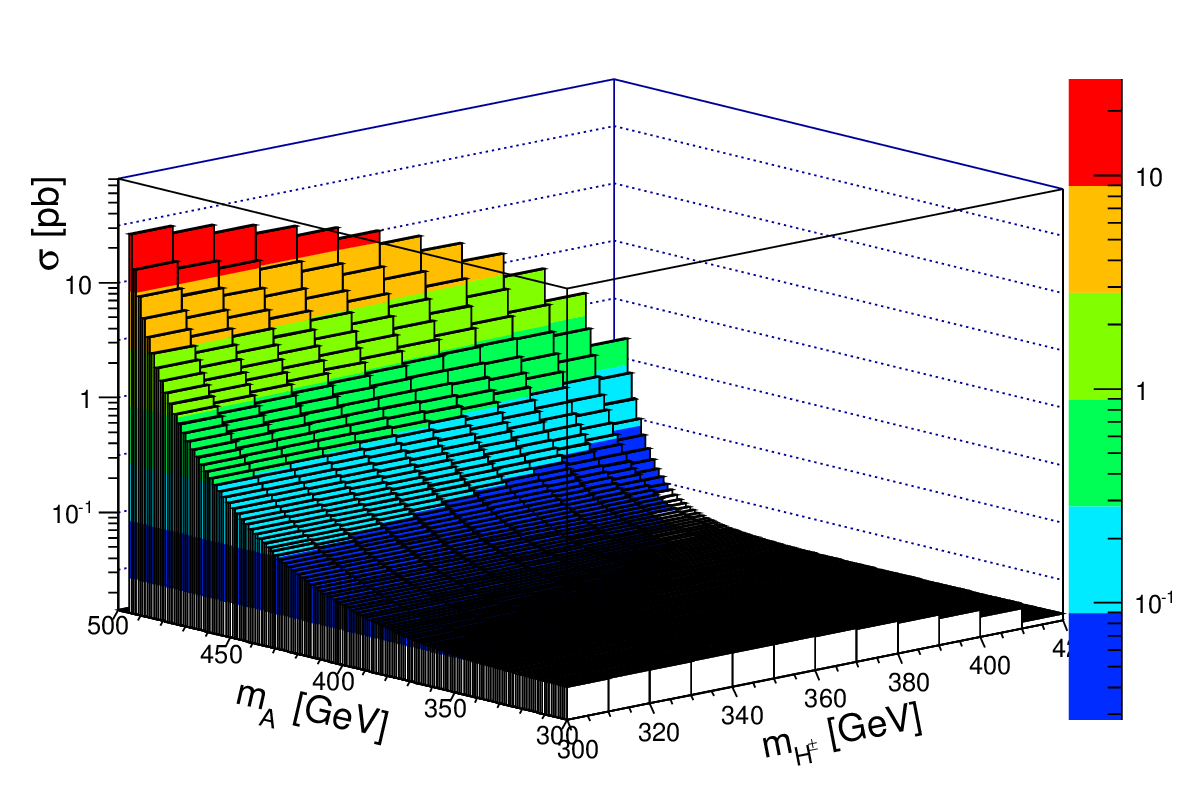}
\end{center}
\caption{The signal cross section in 2HDM type II as a function of $m_{H^{\pm}}$ and $m_A$ with \tanb= 50.}
\label{xsec50}
\end{figure}

In this section, results of a scan in the Higgs boson mass parameter space are presented. Cross sections are calculated with the use of CompHep package \cite{comphep,comphep2} and the validity of the points in the parameter space in terms of the Higgs potential stability, unitarity and perturbativity and compatibility with collider limits on the Higgs bosons masses is checked using 2HDMC \cite{2hdmc}. This package is also used for the evaluation of $H^{\pm}\ra t\bar{b}$ branching ratios. Although a scan over $\alpha$ values expressed in terms of $s_{\beta-\alpha}$ for fixed values of \tanb, shows no sizable sensitivity to this parameter as shown in Fig. \ref{alphascan}, however, according to the discussion in the previous section, we choose $s_{\beta-\alpha}=0$ for showing results. Moreover as mentioned before, light $h$ and $H$ are assumed, although the mass of the latter does not change the cross section as its propagator, being proportional to $s_{\beta-\alpha}$ is absent in the Feynman diagrams. With these assumptions the parameter space is effectively a sub-space with three free parameters, $M_{H^{\pm}}$, $M_{A}$ and $\tan\beta = \tan\alpha$. Total cross sections are calculated for two values of \tanb= 20 and 50 as shown in Figs. \ref{xsec20} and \ref{xsec50}. 

\begin{figure}
\begin{center}
\includegraphics[width=0.7\textwidth]{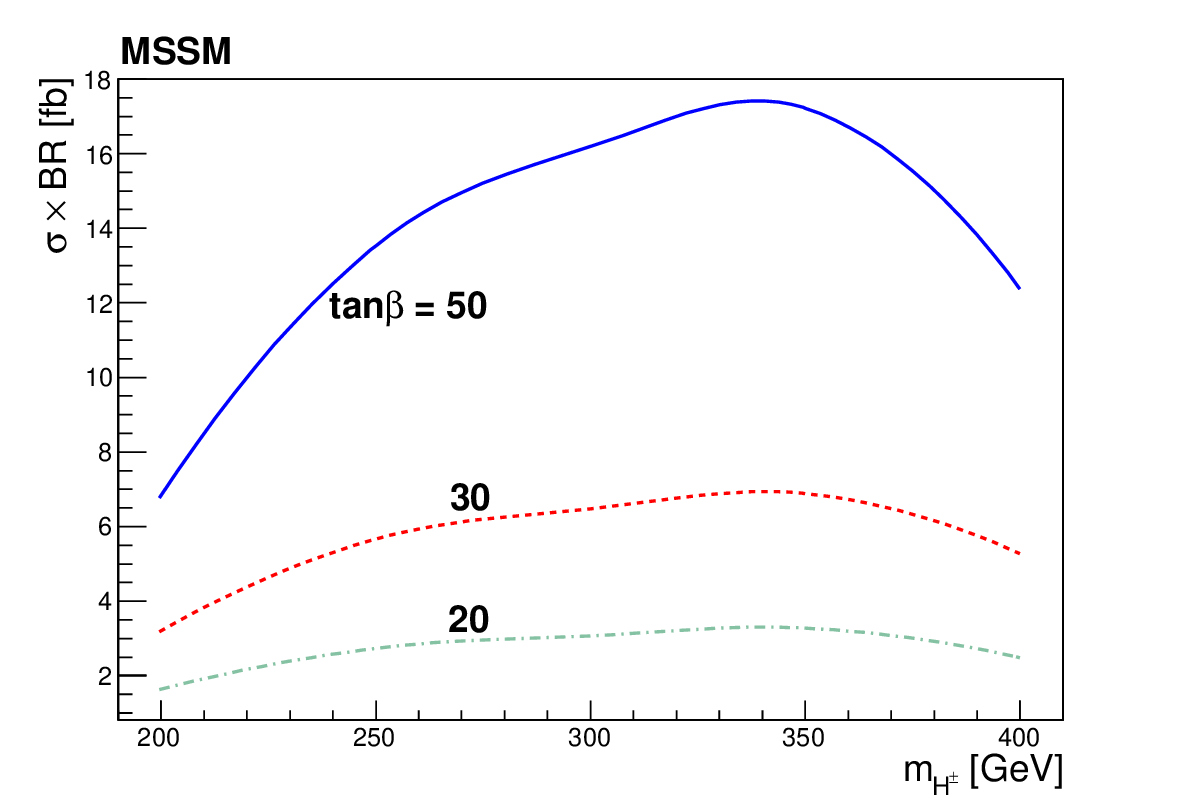}
\end{center}
\caption{The $\sigma \times BR$ as a function of the charged Higgs mass with various \tanb values.}
\label{mssm}
\end{figure}

\begin{figure}
\begin{center}
\includegraphics[width=0.8\textwidth]{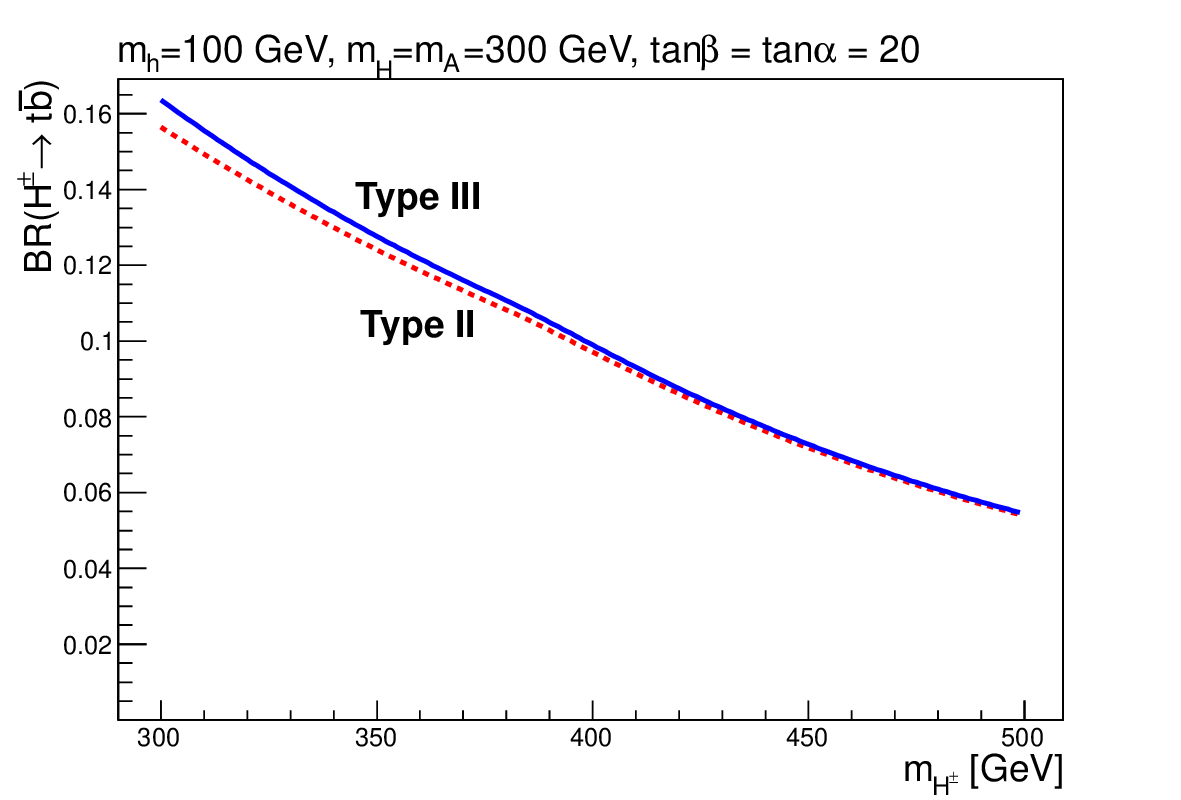}
\end{center}
\caption{The $BR(H^{\pm}\ra t\bar{b})$ as a function of the charged Higgs mass with \tanb= 20.}
\label{br20}
\end{figure}

\begin{figure}
\begin{center}
\includegraphics[width=0.8\textwidth]{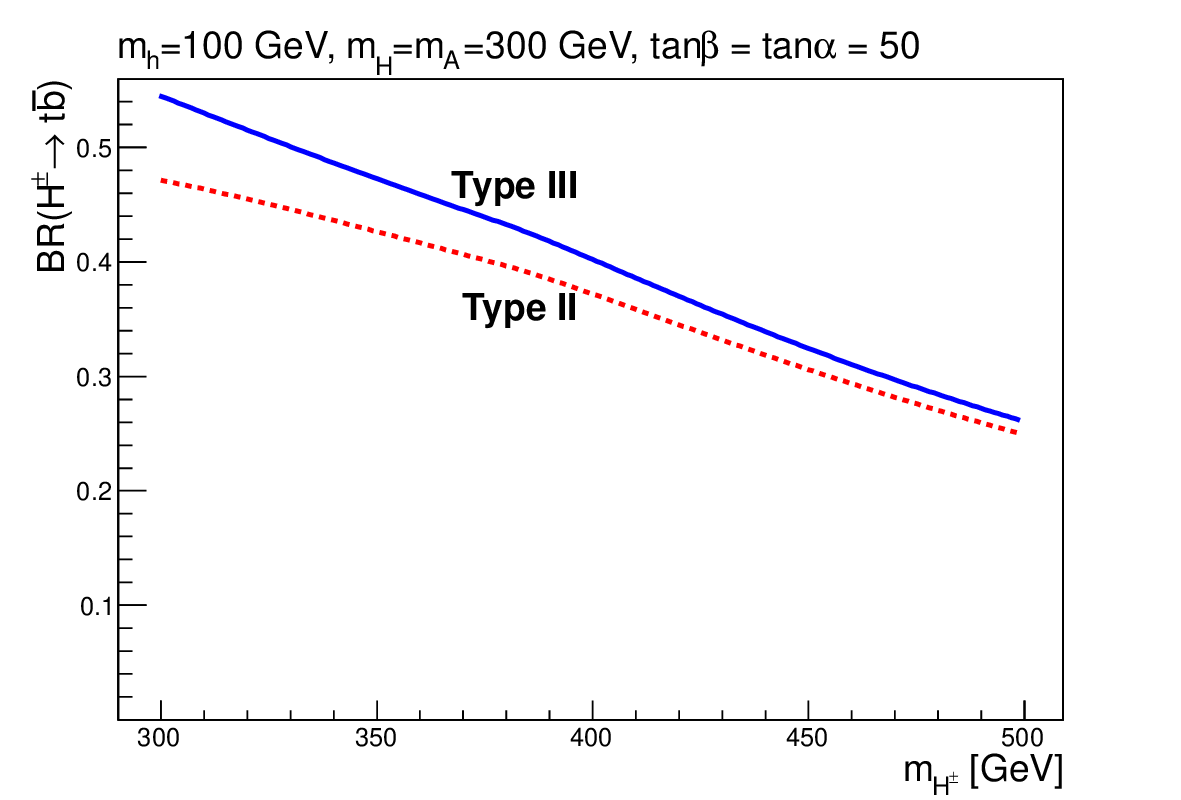}
\end{center}
\caption{The $BR(H^{\pm}\ra t\bar{b})$ as a function of the charged Higgs mass with \tanb= 50.}
\label{br50}
\end{figure}

\begin{figure}
\begin{center}
\includegraphics[width=0.8\textwidth]{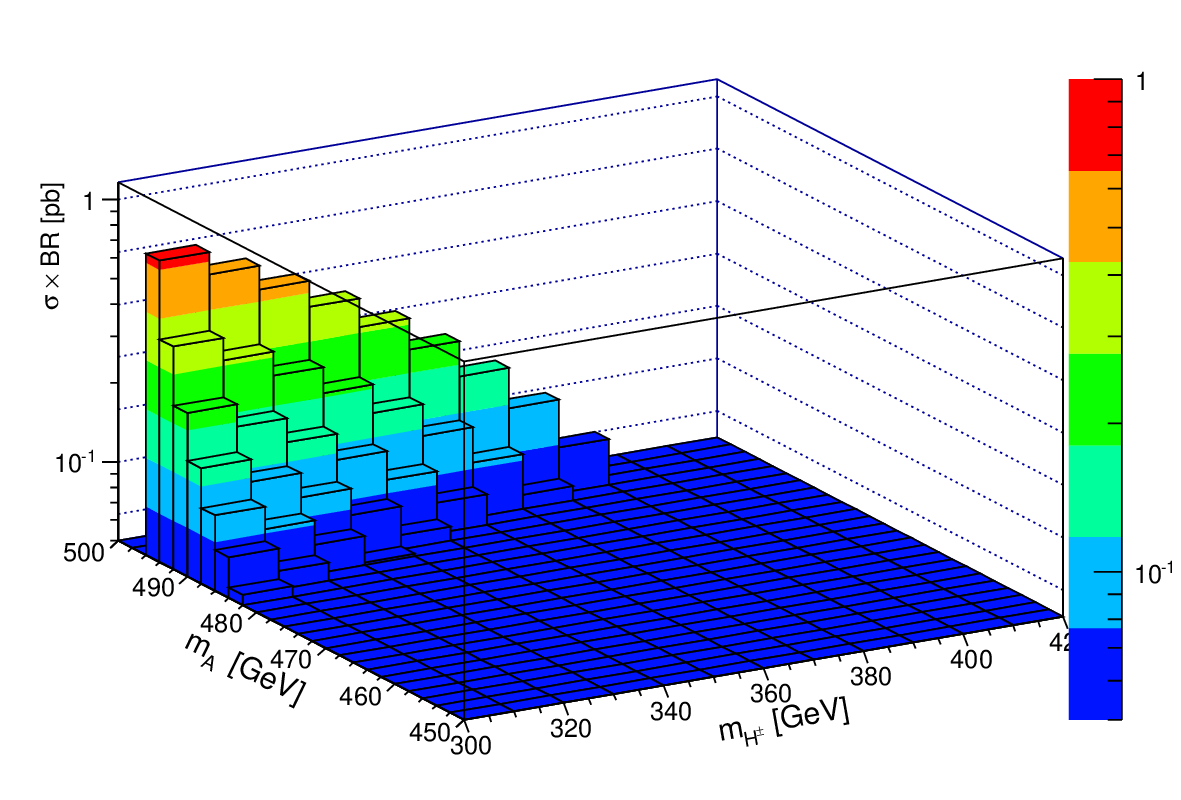}
\end{center}
\caption{The $\sigma \times BR(H^{\pm}\ra t\bar{b})$ with \tanb= 20 as a function of $m_{H^{\pm}}$ and $m_A$ only including points which lead to 5 $\sigma$ signals.}
\label{xsecxbr20}
\end{figure}

\begin{figure}
\begin{center}
\includegraphics[width=0.8\textwidth]{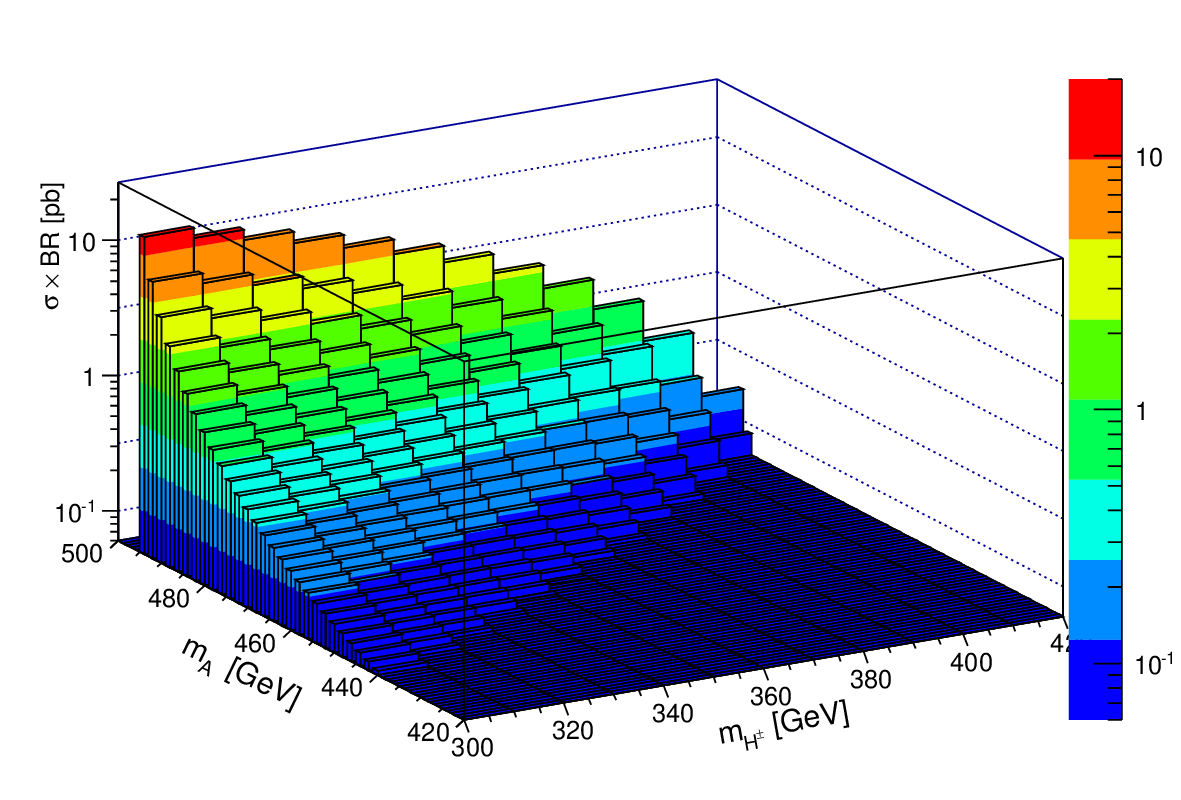}
\end{center}
\caption{The $\sigma \times BR(H^{\pm}\ra t\bar{b})$ with \tanb= 50 as a function of $m_{H^{\pm}}$ and $m_A$ only including points which lead to 5 $\sigma$ signals.}
\label{xsecxbr50}
\end{figure}

As is seen from Figs. \ref{xsec20} and \ref{xsec50}, the total cross section can be enhanced sizably up to several picobarn if a heavy neutral $CP$-odd is present in the model. However the type III 2HDM does not lead to a large cross section as the coupling between $CP$-odd Higgs, $A$ and muons is suppressed by $\cot \beta$. The total cross section in this type is calculated to be $\sim6$ $fb$ and is low to hope for a signal observation. On the contrary in 2HDM type II a 5 $\sigma$ signal at an integrated luminosity of 50 $fb^{-1}$ may be observable. \\
Since the event kinematics with the same final state and particle masses is expected to be the same within MSSM and a 2HDM of the type studied here, results of \cite{mywh} are used as reference. According to \cite{mywh} the MSSM signal significance with \tanb= 50 at 50$fb^{-1}$ is about 1 $\sigma$ (for $m_{H^{\pm}}\simeq 400$ GeV) or better. Therefore the $\sigma \times BR$ for a 5 $\sigma$ signal in 2HDM type II at 50$fb^{-1}$ is obtained by scaling the $\sigma \times BR$ in MSSM by a factor of $\sim 5$ assuming no difference between the background cross sections in MSSM and 2HDM type II. As shown in Fig. \ref{mssm} the $\sigma \times BR$ within MSSM is $\sim 12~fb$ for $m_{H^{\pm}}\simeq 400$ GeV, therefore, at least an amount of $\sigma \times BR\simeq 0.06~pb$ is needed for a 5 $\sigma$ observation in the range 200 GeV $<m_{H^{\pm}}<$ 400 GeV at 50$fb^{-1}$. A similar evaluation leads to the conclusion that with \tanb= 20, $\sigma \times BR\simeq 0.05~pb$ is needed for a 5 $\sigma$ signal. \\
Now the branching ratio of charged Higgs decay to $t\bar{b}$ which is almost independent of $m_A$ is plotted as shown in Figs. \ref{br20} and \ref{br50} for two values of \tanb= 20 and 50. These values are multiplied by the cross section plots in Figs. \ref{xsec20} and \ref{xsec50} respectively. The branching ratios for 2HDM type III have also been illustrated in Figs. \ref{br20} and \ref{br50}, however, although being higher than the corresponding ones from type II, they are not used anymore due to the low cross sections with 2HDM type III. Figures \ref{xsecxbr20} and \ref{xsecxbr50} show the 5 $\sigma$ contours with \tanb= 20 (50) respectively, i.e., the resulting $\sigma \times BR$ values which result in 5 $\sigma$ signals. As an example with \tanb= 20 (50) and $m_{H^{\pm}}=300$ GeV, the charged Higgs signal is observable in a 2HDM type II if $m_{A}>480~(440)$ GeV.
\section{Conclusions}
The charged Higgs production in association with $W^{\pm}$ was studied with a focus on its cross section in a general 2HDM. It was shown that with heavy neutral $CP$-odd Higgs bosons, the cross section exceeds that of MSSM providing the opportunity to observe the signal at a muon collider with an integrated luminosity of 50 $fb^{-1}$. The signal $\sigma \times BR(H^{\pm}\ra t\bar{b})$ is low at a 2HDM type III, however, at a 2HDM type II, the charged Higgs mass range 200 GeV $<m_{H^{\pm}}<$ 400 GeV can be observable already at 50 $fb^{-1}$ if $m_A>480~(440)$ GeV with $\tan\beta= 20~(50)$. Observation of a charged Higgs in the mass range mentioned above with early data of 50 $fb^{-1}$ can be considered as a hint for existence of a heavy neutral Higgs boson which does not fit to the framework of MSSM and its constraints.

\end{document}